\documentclass{yearbook20}
\usepackage{proof}
\usepackage{stmaryrd}
\usepackage{xcolor}
\usepackage{url}

\begin{document}

\LongTitle{Uniqueness of logical connectives in a bilateralist setting}
\ShortTitle{Uniqueness of logical connectives in a bilateralist setting}

\AuthorA{Sara Ayhan}{Ruhr University Bochum, Department of Philosophy I}{Germany}{sara.ayhan@rub.de}
\AuthorAThanks{I would like to thank Heinrich Wansing for the opportunity to discuss this topic and paper extensively and for his feedback, which is always helpful and on point.}



\Abstract{In this paper I will show the problems that are encountered when dealing with uniqueness of connectives in a bilateralist setting within the larger framework of proof-theoretic semantics and suggest a solution.
Therefore, the logic \texttt{2Int} is suitable, for which I introduce a sequent calculus system, displaying - just like the corresponding natural deduction system - a consequence relation for provability as well as one dual to provability.
I will propose a modified characterization of uniqueness incorporating such a duality of consequence relations, with which we can maintain uniqueness in a bilateralist setting.}


\MakeTitlePage

\section{Introduction}

\indent 
The question of uniqueness is the question whether a connective is characterized by the rules governing its use in a way that there is \emph{at most} one connective playing its specific inferential role.
The usual way to test this is to create a `copy-cat' connective governed by the same rules and show that formulas containing these connectives are interderivable.
Important work has been conducted showing the problematic features of certain logics leading to the failure of uniqueness for some connectives in these systems, along with refinements of the requirements for uniqueness (cf. section 3.2).  
In this paper I will deal with bilateralist proof systems, more specifically with proof systems for the logic \texttt{2Int}, which are bilateral in that they display two consequence relations: one for provability and one for dual provability (cf. section 2.2).
In such a setting, according to the common understanding of uniqueness, the question could be raised, whether this bilateralist proof-theoretic semantics (PTS) framework does not lead to different meanings depending on whether we prove or refute.
Making this problem and my solution fully understandable requires laying some groundwork on bilateralism (cf. section 2.1) and uniqueness (cf. section 3.1) first. 
My aim is to show that the problems occurring in a bilateralist setting extend the problematic settings and solutions to ensure uniqueness that have been detected so far.
Finally, I will propose a modification of our characterization of uniqueness that enables us to deal with uniqueness in bilateralism (cf. section 3.3).

\section{Bilateralism}
\subsection{Bilateralism and proof-theoretic semantics}
The topic of bilateralism has received more and more attention in different areas within the past years including the area of PTS.
In a nutshell, bilateralism is the view that dual concepts like truth and falsity, assertion and denial, or, in our context, proof and refutation should each be considered equally important, and not, like it is traditionally done, to concentrate solely on the former concepts.
The debate started out in the context of considerations regarding an approach to the meaning of logical connectives, called “proof-theoretic semantics” (cf. \citep{sep-proof-theoretic-semantics}, for an extensive overview of this area, as well as \citep{Francez2}, which also covers the relation to bilateralism).
In PTS, situated in the broader context of \emph{inferentialism}, the meaning of logical connectives is determined by the rules of inference that govern their use in proofs.
Bilateralism is therefore an approach to meaning which questions the established view, famously held by \citet{Frege} and especially endorsed by Dummett (e.g. in \citeyearpar{Dummett1976, Dummett1981, Dummett1991}), that denying a proposition $A$ is equal to asserting the negation of $A$.\footnote{For an analysis of the established view as well as different ways to tackle it cf. also \citep{Ripley}.}
This has been opposed by several authors claiming that denial is a concept prior to negation and hence, should not be analysed in terms of it (cf. \citealp{Martin, Restall}).
Thus, bilateralism demands an equal consideration of these dual concepts in that they should \textit{both} be taken as primitive concepts, i.e. not reducible to each other.

Applying this to the proof-theoretic context, this amounts to demanding a proof system not only to characterize the proof (or verification) conditions of connectives but also their refutation (or falsification) conditions.
Traditionally, in proof systems like natural deduction systems, the focus is only on the former, whereas, if we consider these notions to be on a par, we need to extend these systems with rules that capture falsification conditions.
This is what Rumfitt \citeyearpar{Rumfitt} proposes in his seminal paper connecting bilateralism and PTS, in which he introduces a natural deduction system with signed formulas for assertion and denial.
\citet{Wansing2017} goes one step further and argues that considering the speech acts of assertion and denial as well as their internally corresponding attitudes of judgment and dual judgment on a par, gives rise to also considering a consequence relation dual to our usual consequence relation. 
He claims that, in order to take bilateralism seriously in the context of proof theory, we need to embed this principle of duality on a level deeper than that of formulas: Next to our usual consequence relation ($\vdash^{+}$), which captures the notion of verification from premises to conclusion, we also need to consider a \emph{dual} consequence relation ($\vdash^{-}$) capturing the dual notion of falsification from premises to conclusion.\footnote{In the spirit of Hacking's \citeyearpar[p. 292]{Hacking} conception of the sequent calculus as a metatheory, I use ``$\vdash^{+}$'' and ``$\vdash^{-}$'' both when talking about consequence relations in the metalanguage as well as for the sequent signs in the sequent calculus system which I will introduce below.}

\subsection{Bilateralist calculi: \texttt{N2Int} and \texttt{SC2Int}}
Therefore, \citet{Wansing2017} devises a natural deduction system for the bi-intuitionistic logic \texttt{2Int}, which comprises not only proofs (indicated by using single lines) but also \emph{dual proofs} (indicated by using double lines).
 Also, a distinction is drawn in the premises between \emph{assumptions} (taken to be verified) and \emph{counterassumptions} (taken to be falsified).
This is indicated by an ordered pair $(\Gamma; \Delta)$ (with $\Gamma$ and $\Delta$ being finite, possibly empty multisets) of assumptions ($\Gamma$) and counterassumptions ($\Delta$).
Single square brackets denote a possible discharge of assumptions, while double square brackets denote a possible discharge of counterassumptions.
The language $\mathcal{L}_{2Int}$ of \texttt{2Int}, as given by Wansing, is defined in Backus-Naur form as follows:\\
$A ::= p \mid \bot \mid \top \mid (A \wedge A) \mid (A \vee A) \mid (A \rightarrow A) \mid (A \Yleft A)$.\\
I will in general use $p, q, r,...$ for atomic formulas, $A, B, C,...$ for arbitrary formulas, and $\Gamma, \Delta, \Gamma',...$ for multisets of formulas.
In a rule, the formula containing the respective connective of that rule is called the \emph{principal formula}, while its components mentioned explicitly in the premises are called the \emph{active formulas}.

As can be seen, we have a non-standard connective in this language, namely the operator of co-implication $\Yleft$,\footnote{Sometimes also called ``pseudo-difference'', e.g. in \citep{Rauszer}, or ``subtraction'', e.g. in \citep{Restall1997}, and used with different symbols.} which acts as a dual to implication, just like conjunction and disjunction can be seen as dual connectives.
With that we are in the realms of so-called bi-intuitionistic logic, which is a conservative extension of intuitionistic logic by co-implication.
Note that there is also a use of ``bi-intuitionistic logic'' in the literature to refer to a specific system, namely \texttt{BiInt}, also called ``Heyting-Brouwer logic''.
Co-implication is there to be understood to internalize the preservation of non-truth from the conclusion to the premises in a valid inference. 
The system \texttt{2Int}, which is treated here, uses the same language as \texttt{BiInt}, but the meaning of co-implication differs in that it internalizes the preservation of falsity from the premises to the conclusion in a dually valid inference \citep[cf.][p. 30ff.]{Wansing2016a, Wansing2016b, Wansing2017}.

From the viewpoint of bilateralism, i.e. considering falsificationism being on a par with verificationism, it is quite natural to extend our language by a connective for co-implication.
The reason for this is that co-implication plays the same role in falsificationism as implication in verificationism: Both can be understood to express a concept of entailment in the object language.
If we expect $\vdash^{+}$ to capture verification from the premises to the conclusion in a valid inference and $\vdash^{-}$ to capture falsification from the premises to the conclusion in a dually valid inference, then, just like implication internalizes provability in that we have in our system $(A; \emptyset) \vdash^{+} B$ iff $(\emptyset; \emptyset) \vdash^{+} A \rightarrow B$, likewise co-implication internalizes dual provability in that we have $(\emptyset; A) \vdash^{-} B$ iff $(\emptyset; \emptyset) \vdash^{-} B \Yleft A$.

With the two implication connectives also two negation connectives are defined: intuitionistic negation with $\neg A$ := $A \rightarrow \bot$ and co-negation with $-A$ := $\top \Yleft A$.
Concerning switching between proofs and dual proofs, there is a division of labour between those negations in that we can move from proofs to dual proofs with intuitionistic negation and from dual proofs to proofs with co-negation: $(\Gamma; \Delta) \vdash^{+} A$ iff $(\Gamma; \Delta) \vdash^{-} \neg A$ and $(\Gamma; \Delta) \vdash^{-} A$ iff $(\Gamma; \Delta) \vdash^{+} - A$.\footnote{I will not consider negation further in this paper, since I am concerned with connectives which are defined by their rules. Cf. \citep{Wansing2016a, Wansing2016b, Wansing2017} for a more detailed discussion, though.}

Besides the usual introduction and elimination rules (henceforth: the \emph{proof rules}) for intuitionistic logic, the natural deduction system \texttt{N2Int}, which is presented below, also contains rules that allow us to introduce and eliminate our connectives into and from dual proofs.
These so-called \emph{dual proof rules} are obtained by a dualization of the proof rules (cf. \citealp[p. 32-34]{Wansing2017}, for the description and the rules of the calculus) and having these two independent sets of rules is exactly what reflects the bilateralism of the proof system.

\vspace{0.3cm}
\textbf{\texttt{N2Int}}
\vspace{0.2cm}

{\footnotesize
\quad \hspace{-0.8cm}  
\infer[\scriptstyle\bot E]{A}{\infer{\bot}{\infer*{}{(\Gamma;\Delta)}}}
\quad \hspace{-0.35cm}  
\infer=[\scriptstyle\top E^{d}]{A}{
	\infer={\top}{\infer*{}{(\Gamma;\Delta)}}}
\quad \hspace{0.8cm}  
\infer[\scriptstyle\wedge I]{A \wedge B}{\;\infer{A}{\infer*{}{(\Gamma ; \Delta)}} \quad \quad \infer{B}{\infer*{}{(\Gamma';\Delta')}}}
\quad \hspace{-0.35cm}  
\infer[\scriptstyle\wedge E_{1}]{A}{\infer{A \wedge B}{\infer*{}{(\Gamma;\Delta)}}}
\quad \hspace{-0.35cm}  
\infer[\scriptstyle\wedge E_{2}]{B}{\infer{A \wedge B}{\infer*{}{(\Gamma;\Delta)}}}

\vspace{0.2cm}
\quad \hspace{-0.8cm}  
\infer=[\scriptstyle\wedge I^{d}_{1}]{A \wedge B}{\infer={A}{\infer*{}{(\Gamma;\Delta)}}}
\quad \quad
\infer=[\scriptstyle\wedge I^{d}_{2}]{A \wedge B}{\infer={B}{\infer*{}{(\Gamma;\Delta)}}}
\quad
\hspace{-0.5cm}  
\infer=[\scriptstyle\wedge E^{d}]{C}{\;\;\;\;\;\;\;\;\;\infer={A \wedge B}{\infer*{}{(\Gamma;\Delta)}} \quad \infer={C}{\infer*{}{(\Gamma';\Delta', \llbracket A\rrbracket)}} \quad \infer={C}{\infer*{}{(\Gamma''; \Delta'', \llbracket B\rrbracket)}}}

\vspace{0.2cm}
\quad  \hspace{-0.8cm}  
\infer[\scriptstyle\vee I_{1}]{A \vee B}{\infer{A}{\infer*{}{(\Gamma;\Delta)}}}
\quad  \quad
\infer[\scriptstyle\vee I_{2}]{A \vee B}{\infer{B}{\infer*{}{(\Gamma;\Delta)}}}
\quad 
\hspace{-0.5cm} 
\infer[\scriptstyle\vee E]{C}{\;\;\;\;\;\;\;\;\;\infer{A \vee B}{\infer*{}{(\Gamma;\Delta)}} \quad \infer{C}{\infer*{}{(\lbrack A\rbrack, \Gamma'; \Delta')}} \quad \infer{C}{\infer*{}{(\lbrack B \rbrack, \Gamma'';\Delta'')}}}

\vspace{0.2cm}

\quad \hspace{-0.8cm} 
\infer=[\scriptstyle\vee I^{d}]{A \vee B}{\;\infer={A}{\infer*{}{(\Gamma;\Delta)}}\quad \quad\infer={B}{\infer*{}{(\Gamma';\Delta')}} }
\quad  \quad \quad
\infer=[\scriptstyle\vee E^{d}_{1}]{A}{\infer={A \vee B}{\infer*{}{(\Gamma;\Delta)}}}
\quad  \quad
\infer=[\scriptstyle\vee E^{d}_{2}]{B}{\infer={A \vee B}{\infer*{}{(\Gamma;\Delta)}}}

\vspace{0.2cm}

\quad  \hspace{-0.8cm}
\infer[\scriptstyle\rightarrow I]{A \rightarrow B}{
	\infer{B}{\infer*{}{([A],\Gamma;\Delta)}}}
\quad \hspace{-0.35cm} 
\infer[\scriptstyle\rightarrow E]{B}{\infer{A}{\infer*{}{(\Gamma;\Delta)}} \ \quad \infer{A \rightarrow B}{\infer*{}{(\Gamma';\Delta')}}}
\quad \hspace{-0.35cm}
\infer=[\scriptstyle\rightarrow I^{d}]{A \rightarrow B}{\;\infer{A}{\infer*{}{(\Gamma;\Delta)}}\quad \infer={B}{\infer*{}{(\Gamma';\Delta')}}}
\quad \hspace{-0.35cm}
\infer[\scriptstyle\rightarrow E^{d}_{1}]{A}{
	\infer={A \rightarrow B}{\infer*{}{(\Gamma;\Delta)}}}
\quad \hspace{-0.35cm}
\infer=[\scriptstyle\rightarrow E^{d}_{2}]{B}{
	\infer={A \rightarrow B}{\infer*{}{(\Gamma;\Delta)}}}

\vspace{0.2cm}

\quad \hspace{-0.8cm} 
\infer[\scriptstyle\Yleft I]{A \Yleft B}{\;\infer{A}{\infer*{}{(\Gamma;\Delta)}}\quad \quad\infer={B}{\infer*{}{(\Gamma';\Delta')}}}
\quad \hspace{-0.35cm} 
\infer[\scriptstyle\Yleft E_{1}]{A}{
	\infer{A \Yleft B}{\infer*{}{(\Gamma;\Delta)}}}
\quad \hspace{-0.35cm} 
\infer=[\scriptstyle\Yleft E_{2}]{B}{
	\infer{A \Yleft B}{\infer*{}{(\Gamma;\Delta)}}}
\quad  \hspace{-0.35cm} 
\infer=[\scriptstyle\Yleft I^{d}]{B \Yleft A}{
	\infer={B}{\infer*{}{(\Gamma; \Delta, \llbracket A \rrbracket)}}}
\quad \hspace{-0.35cm} 
\infer=[\scriptstyle\Yleft E^{d}]{B}{\;\;\;\;\infer={B \Yleft A}{\infer*{}{(\Gamma;\Delta)}} \quad \quad \infer={A}{\infer*{}{(\Gamma';\Delta')}}}}
\vspace{0.3cm}

What I will present here additionally, is a sequent calculus, which I will call \texttt{SC2Int}.
\texttt{SC2Int} corresponds to \texttt{N2Int} in that we have a proof in \texttt{N2Int} of $A$ from the pair $(\Gamma; \Delta)$, iff the sequent $(\Gamma; \Delta) \vdash^{+} A$ is derivable in \texttt{SC2Int} and we have a dual proof of $A$ from the pair $(\Gamma; \Delta)$, iff the sequent $(\Gamma; \Delta) \vdash^{-} A$ is derivable in \texttt{SC2Int}.
While \citet{Wansing2017} proves a normal form theorem for \texttt{N2Int}, for \texttt{SC2Int} also a cut-elimination theorem can be proven \citep{Ayhan}. 
Since this means that our system enjoys the subformula property, this ensures the conservativeness of our system.\footnote{The exact relation between conservativeness and cut-elimination is debatable and, more specifically, depends on the system that is used \citep[cf.][]{Hacking, Kremer} but given that we can also prove admissibility of the other structural rules, this should be a safe assumption for our system.}
Sequents are of the form $(\Gamma; \Delta) \vdash^{*} C$ (with $\Gamma$ and $\Delta$ being finite, possibly empty multisets), which are read as ``From the verification of all formulas in $\Gamma$ and the falsification of all formulas in $\Delta$ one can derive the verification (resp. falsification) of $C$ for $*= +$ (resp. $* = -$)''.  
Within the right introduction rules we need to distinguish whether the derivability relation expresses verification or falsification by using the superscripts + and -.
Within the left rules this is not necessary, but what is needed here instead is distinguishing an introduction of the principal formula into the \textit{assumptions} (indexed by superscript $a$) from an introduction into the \textit{counterassumptions} (indexed by superscript $c$).
Thus, the set of \emph{proof rules} in \texttt{SC2Int} consists of the rules marked with + or with $a$, while the set of \emph{dual proof rules} consists of the rules marked with - or with $c$.
When a rule contains multiple occurrences of $*$, application of this rule requires that all such occurrences are instantiated in the same way, i.e. either as + or as -. \\

\textbf{\texttt{SC2Int}}

{\footnotesize
\vspace{0.1cm}
\quad \hspace{-0.8cm}  For $\ast \in \{+, -\}:
$
\vspace{0.1cm}

\quad \hspace{-0.8cm}  
\infer[\scriptstyle Rf^{+}]{(\Gamma, p; \Delta) \vdash^{+} p}{}
\quad \quad
\infer[\scriptstyle Rf^{-}]{(\Gamma; \Delta, p) \vdash^{-} p}{}

\vspace{0.1cm}
\quad \hspace{-0.8cm}
\infer[\scriptstyle\bot R^{-}]{(\Gamma; \Delta) \vdash^{-} \bot}{}
\quad \hspace{-0.35cm}  
\infer[\scriptstyle\bot L^{a}]{(\Gamma, \bot; \Delta) \vdash^{*} C}{}
\quad \hspace{-0.35cm}
\infer[\scriptstyle\top R^{+}]{(\Gamma; \Delta) \vdash^{+} \top}{}
\quad \hspace{-0.35cm}
\infer[\scriptstyle\top L^{c}]{(\Gamma; \Delta, \top) \vdash^{*} C}{}

\vspace{0.3cm}
\quad  \hspace{-0.8cm} 
\infer[\scriptstyle\wedge R^{+}]{(\Gamma; \Delta) \vdash^{+} A \wedge B}{(\Gamma; \Delta) \vdash^{+} A \quad \quad (\Gamma; \Delta) \vdash^{+} B}
\quad  \quad
\infer[\scriptstyle\wedge L^{a}]{(\Gamma, A \wedge B; \Delta) \vdash^{*} C}{(\Gamma, A, B; \Delta) \vdash^{*} C}

\vspace{0.1cm}
\quad \hspace{-0.8cm}  
\infer[\scriptstyle\wedge R^{-}_{1}]{(\Gamma; \Delta) \vdash^{-} A \wedge B}{(\Gamma; \Delta) \vdash^{-} A}
\quad \hspace{-0.35cm}
\infer[\scriptstyle\wedge R^{-}_{2}]{(\Gamma; \Delta) \vdash^{-} A \wedge B}{(\Gamma; \Delta) \vdash^{-} B}
\quad \hspace{-0.35cm}
\infer[\scriptstyle\wedge L^{c}]{(\Gamma; \Delta, A \wedge B) \vdash^{*} C}{(\Gamma; \Delta, A) \vdash^{*} C \quad (\Gamma; \Delta, B) \vdash^{*} C}

\vspace{0.3cm}
\quad \hspace{-0.8cm} 
\infer[\scriptstyle\vee R^{+}_{1}]{(\Gamma; \Delta) \vdash^{+} A \vee B}{(\Gamma; \Delta) \vdash^{+} A}
\quad \hspace{-0.35cm}
\infer[\scriptstyle\vee R^{+}_{2}]{(\Gamma; \Delta) \vdash^{+} A \vee B}{(\Gamma; \Delta) \vdash^{+} B}
\quad  \hspace{-0.35cm}
\infer[\scriptstyle\vee L^{a}]{(\Gamma, A \vee B; \Delta) \vdash^{*} C}{(\Gamma, A; \Delta) \vdash^{*} C \quad (\Gamma, B; \Delta) \vdash^{*} C}

\vspace{0.1cm}
\quad \hspace{-0.8cm} 
\infer[\scriptstyle\vee R^{-}]{(\Gamma; \Delta) \vdash^{-} A \vee B}{(\Gamma; \Delta) \vdash^{-} A \quad \quad (\Gamma; \Delta) \vdash^{-} B}
\quad  \quad
\infer[\scriptstyle\vee L^{c}]{(\Gamma; \Delta, A \vee B) \vdash^{*} C}{(\Gamma; \Delta, A, B) \vdash^{*} C}

\vspace{0.3cm}
\quad \hspace{-0.8cm}  
\infer[\scriptstyle\rightarrow R^{+}]{(\Gamma; \Delta) \vdash^{+} A \rightarrow B}{(\Gamma, A; \Delta) \vdash^{+} B}
\quad \quad
\infer[\scriptstyle\rightarrow L^{a}]{(\Gamma, A \rightarrow B; \Delta) \vdash^{*} C}{(\Gamma, A \rightarrow B; \Delta) \vdash^{+} A \quad \quad (\Gamma, B; \Delta) \vdash^{*} C}

\vspace{0.1cm}
\quad  \hspace{-0.8cm} 
\infer[\scriptstyle\rightarrow R^{-}]{(\Gamma;\Delta) \vdash^{-} A \rightarrow B}{(\Gamma;\Delta)\vdash^{+} A \quad \quad (\Gamma;\Delta) \vdash^{-} B}
\quad \quad
\infer[\scriptstyle\rightarrow L^{c}]{(\Gamma; \Delta, A \rightarrow B) \vdash^{*} C}{(\Gamma, A; \Delta, B) \vdash^{*} C}

\vspace{0.3cm}
\quad \hspace{-0.8cm}  
\infer[\scriptstyle\Yleft R^{+}]{(\Gamma; \Delta) \vdash^{+} A \Yleft B}{(\Gamma; \Delta) \vdash^{+} A \quad \quad (\Gamma; \Delta) \vdash^{-} B}
\quad \quad
\infer[\scriptstyle\Yleft L^{a}]{(\Gamma, A \Yleft B;\Delta) \vdash^{*} C}{(\Gamma, A; \Delta, B) \vdash^{*} C}

\vspace{0.1cm}
\quad  \hspace{-0.8cm}
\infer[\scriptstyle\Yleft R^{-}]{(\Gamma; \Delta) \vdash^{-} A \Yleft B}{(\Gamma; \Delta, B) \vdash^{-} A}
\quad \quad
\infer[\scriptstyle\Yleft L^{c}]{(\Gamma; \Delta, A \Yleft B) \vdash^{*} C}{(\Gamma; \Delta, A \Yleft B) \vdash^{-} B \quad \quad (\Gamma; \Delta, A) \vdash^{*} C}}

\vspace{0.3cm}
The following structural rules of weakening, contraction, and cut can be shown to be admissible in \texttt{SC2Int}:

{\footnotesize
\vspace{0.3cm}
\quad  \hspace{-0.8cm}
\infer[\scriptstyle W^{a}]{(\Gamma, A; \Delta) \vdash^{*} C}{(\Gamma;\Delta) \vdash^{*} C}
\quad \hspace{-0.4cm}
\infer[\scriptstyle W^{c}]{(\Gamma;\Delta, A) \vdash^{*} C}{(\Gamma;\Delta) \vdash^{*} C}
\quad \hspace{-0.4cm}  
\infer[\scriptstyle C^{a}]{(\Gamma, A;\Delta) \vdash^{*} C}{(\Gamma, A, A; \Delta) \vdash^{*} C}
\quad \hspace{-0.4cm}
\infer[\scriptstyle C^{c}]{(\Gamma; \Delta, A) \vdash^{*} C}{(\Gamma;\Delta, A, A) \vdash^{*} C}

\vspace{0.3cm}
\quad  \hspace{-0.8cm}
\infer[\scriptstyle Cut^{a}]{(\Gamma, \Gamma'; \Delta, \Delta') \vdash^{*} C}{(\Gamma;\Delta) \vdash^{+} D \quad \quad (\Gamma', D; \Delta') \vdash^{\ast} C}
\quad 
\infer[\scriptstyle Cut^{c}]{(\Gamma, \Gamma'; \Delta, \Delta') \vdash^{*} C}{(\Gamma;\Delta) \vdash^{-} D \quad \quad (\Gamma'; \Delta', D) \vdash^{*} C}}

\section{Uniqueness}
\subsection{The notion of uniqueness}
The issue of uniqueness has not received much attention in the literature.
It was introduced more or less \emph{en passant} in Belnap's \citeyearpar{Belnap} famous response to the \texttt{tonk}-attack by \citet{Prior} against an inferentialist view on the meaning of connectives.\footnote{Belnap refers to a lecture by Hi\.{z} as being the actual origin of this idea.}
Prior's intention in using \texttt{tonk} is to show that it leads the idea of PTS\footnote{The term ``proof-theoretic semantics'' emerged much later of course but I use it whenever the idea fits to whatever terminology may be used in other places.} \emph{ad absurdum}.
He argues that if the rules of inference governing the use of a connective would indeed be all there is to the meaning of it, then nothing would prevent the inclusion of a seemingly non-sensical connective, which ultimately trivializes our system, since it allows anything to be derived from everything.
Belnap's proposal to solve this so-called \emph{existence} issue of connectives was to demand extensions of a given system to be ``conservative''.
In addition to that, he claims, one could wonder about the \emph{uniqueness} issue of connectives.
Once we have settled that it is allowed to extend our system with a certain connective, we can ask whether the rules of inference governing the connective characterize this connective \emph{uniquely}.

Uniqueness as a requirement for a connective means that characterizing its inference rules amounts to exactly specifying its role in inference.
There can be \emph{at most} one connective playing this role; duplication of that connective with the same characterizing rules does not change its behaviour, neither in the premises nor in the conclusion.
However, since Belnap's first requirement of conservativeness of the system was seen (by the responding literature and also by himself) to be far more important, the uniqueness requirement was more or less forgotten until it resurfaced in \citep{Do/S-H1985, Do/S-H1988}, which cover quite technical treatments of the issue as well as of connections to other proof-theoretic features.
After that, the topic is absent from the debate for a long time again.
A recent resuming of it can be found in \citep{Naibo}, which targets the question whether the uniqueness condition for connectives is the same as Hacking’s ``deducibility of identicals''-criterion\footnote{The condition that the structural rule of reflexivity for arbitrary formulas is provably admissible for every connective, i.e. each derivation using an application of it with a complex formula can be replaced by a derivation using applications of the rule with only atomic formulas \citep{Hacking}.}. 
\citet{Humberstone2011, Humberstone2019, sep-connectives-logic} is one of the few scholars who treats the topic quite extensively, dedicating one chapter of his monumental work on connectives to the question of uniqueness.
His observations on the connections between (failure of) uniqueness of connectives, proof systems, and features of the consequence relation are of particular importance for the present purpose. 

On the usual account of uniqueness two connectives \# and \#', which are defined by exactly the same set of inference rules and $\vdash$ being the consequence relation generated by the combined set of the rules, play exactly the same inferential role iff it can be shown for all $A$ and $B$ that $A~ \# ~B$ $\dashv \vdash$ $A~ \#' ~B$.  
Let us assume, for the moment, a common intuitionistic calculus and the example of conjunction.
It can easily be shown that $\wedge$ is uniquely characterized by its usual natural deduction (resp. sequent calculus) rules (i.e. in our systems above: by its \emph{proof rules}) governing it, since we can derive $A \wedge B$ from $A \wedge' B$ and vice versa, taking $\wedge'$ to be a connective governed by exactly the same rules as $\wedge$:

{\small
\quad \quad 
	\infer[\scriptstyle\wedge I]{A \wedge B}{\infer[\scriptstyle\wedge' E]{A}{A \wedge' B} \quad \quad \infer[\scriptstyle\wedge' E]{B}{A \wedge' B}}
	\quad
\infer[\scriptstyle\wedge I']{A \wedge' B}{\infer[\scriptstyle\wedge E]{A}{A \wedge B} \quad \quad \infer[\scriptstyle\wedge E]{B}{A \wedge B}}}

\noindent Thus, the interderivability requirement makes clear why, as I mentioned above, it is important to consider the underlying consequence relation when asking about the uniqueness of connectives.

Belnap's \citeyearpar[p. 133]{Belnap} original counterexample for satisfying the uniqueness condition is the connective \texttt{plonk}.
We define \texttt{plonk} by the following rule: $A ~\texttt{plonk}~ B$ can be derived from $B$.  
Since an extension with \texttt{plonk} (in the system Belnap is presupposing) is conservative, it can be stated that there is such a connective.
However, it is not unique, since there can be another connective, which he calls \texttt{plink} defined by exactly the same rule, i.e. $A ~\texttt{plink}~ B$ can be derived from $B$, which can otherwise play a different inferential role.
The uniqueness requirement, as Belnap puts it, demands that another connective specified by exactly the same rules ought to play exactly the same role in inference, both as premise and as conclusion.
In his system with reflexivity, weakening, permutation, contraction, and transitivity as structural rules, this amounts to showing that $A ~\texttt{plonk}~ B$ and $A~\texttt{plink}~B$ are interderivable.
This, however, is not possible given that there is only this one rule governing the connectives and hence, \texttt{plonk} is not uniquely determined by its definition.

\subsection{Problematic settings}

There are several examples of connectives which are not uniquely characterized.
This can be shown not only for `ad hoc' connectives, in the sense that they are only thought of for this purpose, but also for connectives existing in calculi actually used, as e.g. $\neg$ in \texttt{FDE} or $\boxempty$ in system \texttt{K}.\footnote{Or for that matter $\boxempty$ in every normal modal logic except for the Post-complete ones \citep[p. 601-605]{Humberstone2011}. Examples of failure of uniqueness are given in \citep{Humberstone2011,Humberstone2019, Humberstone2020, Naibo}.}
Failure of uniqueness can - among other reasons - occur due to the specific formulation of the proof system, non-congruentiality of the logic or impurity of the rules.
\citet[p. 595f.]{Humberstone2011} emphasizes that what does or does not uniquely characterize a given connective is the \emph{set of rules} governing the connective, while sets of rules can be seen as a set of \emph{conditions on consequence relations}.

The usual system Humberstone refers to when showing the non-uniqueness (e.g. of the examples mentioned in the last paragraph) is what he calls ``sequent-to-sequent rules in the framework \textsc{SET-FMLA}'', i.e. sequent rules with a set of formulas on the left side of the sequent operator and exactly one formula on the right. 
He also gives examples, however, where we have uniqueness in one particular formulation of the rules but not in another.
Negation in Minimal Logic, for example, cannot be uniquely characterized by any collection of \textsc{SET-FMLA}-rules, but can be by others, which allow \emph{at most} one formula on the right side of the sequent operator \citep[p. 186]{Humberstone2020}.
Another example would be that disjunction is not uniquely characterized by its classical (or intuitionistic) rules when those are formulated in a zero-premise \textsc{SET-FMLA} system \citep[p. 600]{Humberstone2011}.

Another important issue concerning uniqueness is the question of congruentiality, which can be a property of connectives, consequence relations, or logics (depending on the specific understanding of those concepts).
A logic is congruential, if for all formulas $A$, $B$, $C$, whenever $A$ and $B$ are equivalent insofar as they are interderivable according to a defined consequence relation of the logic, equivalence also holds when we replace $A$ and $B$ in a more complex formula $C$ \citep{Wojcicki}.\footnote{W\'{o}jcicki actually uses the term ``self-extensional'' instead of ``congruential''. The latter is used by \citet[p. 175]{Humberstone2011} for the case of connectives and consequence relations.}
This is closely connected to the notion of synonymy between formulas, since synonymy means that they are not only equivalent but also that replacing one by the other in any complex formula results in equivalent formulas.
In view of (non-)congruentiality \citet[p. 579f.]{Humberstone2011} refines what I described as `the usual account' (which he calls \emph{uniqueness to within equivalence}) in that he claims that \# is uniquely characterized by its set of rules iff every compound formed by that connective is \emph{synonymous} to every compound (with the same components) formed by \#' governed by exactly the same rules as \#, which he calls \emph{uniqueness to within synonymy}.
This distinction coincides in the congruential case, but when the consequence relation is non-congruential, it can make a difference whether we demand the stronger or the weaker notion \citeyearpar[p. 183, 187]{Humberstone2020}.

Another terminological refinement is needed when we have  systems with connectives governed by impure rules, i.e. rules which govern more than one connective. 
In this case, \citet[p. 580f.]{Humberstone2011} argues, we need to speak of the connective in question being uniquely characterized \emph{in terms of} whichever connective also appears in its rules.
An example would be a connective from another non-congruential logic, namely Nelson's constructive logic with strong negation, \texttt{N4}.\footnote{I choose this example because \texttt{N4} and \texttt{2Int} are related in that strong negation $\sim$ in Nelson's logic can be read as a direct toggle between proofs and dual proofs, if it were added to \texttt{2Int}, i.e. we would have $\vdash^{+} A$ iff $\vdash^{-} \sim A$ and $\vdash^{-} A$ iff $\vdash^{+} \sim A$.} 
The rules governing strong negation, $\sim$, are impure because they also display other connectives, like conjunction and implication.\footnote{At least this is the case for the traditional (unilateral) calculi given for \texttt{N4} \citep[e.g.][p. 97]{Prawitz1965}. In \citep{Kamide}, however, there is a bilateral sequent calculus given for \texttt{N4}, which consists of pure rules only. This can be achieved, as in the case of \texttt{SC2Int}, with a system expressing different consequence relations. Likewise, \citet{Drobyshevich} introduces the notion of a signed consequence relation between a set of signed formulas and a single signed formula as a bilateral variant of the notion of a Tarskian consequence relation and gives a bilateral natural deduction system for \texttt{N4}, which also contains pure rules only.}
Thus, if we would ask for the uniqueness of $\sim$ in \texttt{N4} (with impure rules), the question would always have to be ``Is $\sim$ uniquely characterized by its rules in terms of $\wedge$ and $\rightarrow$?''.
In \texttt{N4} this negation leads to the system's non-congruentiality, since for two formulas to be equivalently replaceable \emph{in all contexts} it is not sufficient for the formulas to be provably equivalent, but additionally, we also need equivalence between the \emph{negated} formulas.\footnote{A counterexample to congruentiality of \texttt{N4} is that equivalence holds between $\sim(A \rightarrow B)$ and $(A~ \wedge \sim B)$ but not between $\sim\sim(A \rightarrow B)$ and $\sim(A~ \wedge \sim B)$ \citep[p. 445]{Wansing2016a}.}
For uniqueness this would mean that firstly, we would have to demand uniqueness to within synonymy. 
Secondly, it would tie uniqueness in this system to strong negation, since we would have to demand not only the interderivability of all formulas containing the connective in question with the formula containing the `copy-cat' connective, but also the same interderivability with the strongly negated formulas.
However, given that strong negation can only be uniquely characterized in terms of other connectives, this does not seem like a desirable system or a good solution to recover uniqueness.

\subsection{Problems in a bilateralist system}
The problem that occurs when asking about uniqueness in a bilateralist setting is closely connected to the last point addressed.
However, I will show that in this case a much more intuitive solution can be given.\footnote{It would exceed the scope of this paper to consider all kinds of bilateral systems here, but e.g. for Rumfitt's system with signed formulas the same problem would arise in a different guise, although the solution to maintain uniqueness might be - as in \texttt{N4} - not that elegant.}
What causes trouble in the bilateralist proof systems laid out above - if we assume the common characterization of uniqueness (to within equivalence or synonymy) - is that we have two sets of rules for each connective and two consequence relations.
It would make sense then to think of the proof rules as generating the consequence relation for provability and the dual proof rules as generating the dual consequence relation for dual provability.
The specific consequence relation is of course important, since we usually test for uniqueness via interderivability, and in \texttt{2Int} it can be shown for both relations \emph{individually} that our connectives are uniquely characterized by only a part of the whole set of rules. 
Consider the case of conjunction, for example: We can show that $\wedge$ is uniquely characterized by its proof rules, since we can show (cf. derivations in section 3.1) that both $(A \wedge B; \emptyset) \vdash^{+} A \wedge' B$ and $(A \wedge' B; \emptyset) \vdash^{+} A \wedge B$ are derivable.
Likewise, taking $\wedge''$ to be a connective governed by exactly the same $I^{d}$- and $E^{d}$-rules from \texttt{N2Int} as $\wedge$ (resp. $R^{-}$ and $ L^{c}$-rules from \texttt{SC2Int}), we can show that it is also uniquely characterized by its dual proof rules, since $(\emptyset; A \wedge B) \vdash^{-} A \wedge'' B$ and $(\emptyset; A \wedge'' B) \vdash^{-} A \wedge B$ are derivable. 
To show it for \texttt{SC2Int}:

{\tiny
\vspace{0.08cm}
\quad \hspace{-1cm}
\infer[\scriptstyle\wedge L^{c}]{(\emptyset; A \wedge B) \vdash^{-} A \wedge'' B}{\infer[\scriptstyle\wedge'' R^{-}_{1}]{(\emptyset; A) \vdash^{-} A \wedge'' B}{\infer[\scriptstyle\ Rf^{-}]{(\emptyset; A) \vdash^{-} A} {}}\quad \infer[\scriptstyle\wedge'' R^{-}_{2}]{(\emptyset; B) \vdash^{-} A \wedge'' B} {\infer[\scriptstyle\ Rf^{-}]{(\emptyset; B) \vdash^{-} B}{}}}
\quad 
\infer[\scriptstyle\wedge'' L^{c}]{(\emptyset; A \wedge'' B) \vdash^{-} A \wedge B}{\infer[\scriptstyle\wedge R^{-}_{1}]{(\emptyset; A) \vdash^{-} A \wedge B}{\infer[\scriptstyle\ Rf^{-}]{(\emptyset; A) \vdash^{-} A} {}}\quad \infer[\scriptstyle\wedge R^{-}_{2}]{(\emptyset; B) \vdash^{-} A \wedge B} {\infer[\scriptstyle\ Rf^{-}]{(\emptyset; B) \vdash^{-} B}{}}}}

However, there is no possibility to determine by this characterization that there is \emph{only one} connective $\wedge$ because it is not possible to derive the following sequents:

{\footnotesize
\hspace{1cm} $(A \wedge B; \emptyset) \vdash^{+} A \wedge'' B$
\quad \quad
\hspace{1cm} $(\emptyset; A \wedge B) \vdash^{-} A \wedge' B$

\hspace{1cm} $(A \wedge'' B; \emptyset) \vdash^{+} A \wedge B$
\quad\quad
\hspace{1cm} $(\emptyset; A \wedge' B) \vdash^{-} A \wedge B$} 

The difference to \texttt{plonk} and \texttt{plink} is that in this case the one rule governing those connectives was `not enough' to uniquely characterize a role in inference, while here a partial duplication of the rules (with proof rules only or dual proof rules only) is already enough for a unique characterization.
So, in a way, we could say, the bilateral sets of rules overdetermine our connectives.
However, since on the one hand both the proof rules as well as the dual proof rules uniquely characterize a connective, but on the other hand, there is no interderivability `across' the consequence relations possible, how can we know that there is one conjunction with a unique meaning?
Wouldn't that mean that we would be forced to say that there are actually two conjunctions, $\wedge^{+}$ and $\wedge^{-}$, one for the context of provability and one for dual provability?
Thus, we could not confidently claim that our conjunction is uniquely characterized and has only one meaning in a system like \texttt{N2Int} or \texttt{SC2Int}, which would certainly have to be considered problematic.

However, let us take a look at our rules again, especially at the ones for implication and co-implication:
What we can see here is that the different consequence relations are intertwined in characterizing these connectives.
In \texttt{N2Int} this is observable by a mixture of single and double lines in the dual proof rules of implication, $\rightarrow I^{d}$ and $\rightarrow E^{d}_{1}$, and in the proof rules of co-implication, $\Yleft I$ and $\Yleft E_{2}$. 
In \texttt{SC2Int} this is indicated in the dual proof rules of implication, $\rightarrow R^{-}$ and $\rightarrow L^{c}$, as well as in the proof rules of co-implication, $\Yleft R^{+}$ and $\Yleft L^{a}$, by a mixture of $\vdash^{+}$ and $\vdash^{-}$ in the right introduction rules and for the left introduction rules by the fact that active formulas are part of the assumptions \emph{as well as} of the counterassumptions.
Thus, the rules for implication as well as for co-implication need \emph{both} consequence relations in one and the same rule application.
This indicates that it would not be correct to think of the proof rules as generating the consequence relation and the dual proof rules as generating the dual consequence relation.
Instead,  both relations are generated by rules of both sets.\footnote{\texttt{SC2Int} shows this feature of `mixedness' even nicer than \texttt{N2Int}, since in the former we have a $\vdash^{*}$ in \emph{all} left rules, meaning that the rule holds for both verification and falsification.}
And this fact would support the point that we are not allowed to use different duplications of a connective when trying to show its uniqueness.
Thus, when duplicating a connective, we need to use \emph{the same duplication} for both proof rules \emph{and} dual proof rules.
By doing so, it is guaranteed that we are not talking about different connectives in different proof contexts.

So my proposal is to modify our characterization of uniqueness in a way that it also fits the context of bilateralism:
In a bilateralist setting, instead of taking interderivability as a sufficient criterion for uniqueness, we also have to consider dual interderivability.
\begin{description}
\item \textbf{Definition of uniqueness for bilaterally defined connectives:}\\
In a bilateralist setting with consequence relations for verification as well as falsification, two n-place connectives \# and \#', which are defined by exactly the same set of inference rules, play exactly the same inferential role, i.e. are unique, iff for all $A_1, \ldots , A_n$ the formulas $\# (A_1,\ldots, A_n)$ and $\#' (A_1,\ldots, A_n)$ are interderivable as well as dually interderivable. 
To express this formally for the case of \texttt{2Int}:
\begin{itemize}
\item[(i)] $(A~ \# ~B; \emptyset) \vdash^{+} A~ \#' ~B$ and $(A ~\#'~ B; \emptyset) \vdash^{+} A ~\#~ B$
\item[(ii)] $(\emptyset ; A~ \# ~B) \vdash^{-} A~ \#' ~B$ and $(\emptyset ; A ~\#'~ B) \vdash^{-} A ~\#~ B$.
\end{itemize}
 
\end{description}
With this definition of uniqueness we can state that all connectives of \texttt{2Int} are uniquely characterized by their rules with respect to \texttt{N2Int} and \texttt{SC2Int}.\footnote{This also holds for the constants $\top$ and $\bot$, since in the case of n=0, $\# (A_1,\ldots, A_n)$ = $\#$.}

A last question to consider, having Humberstone's distinction in mind, would be if this holds for \emph{uniqueness to within equivalence} only or also for \emph{uniqueness to within synonymy}.
The question needs to be asked since \texttt{2Int} is in fact also a non-congruential logic.
The non-congruentiality in \texttt{2Int} stems from the fact that not all formulas that are equivalent with respect to $\vdash^{+}$ are also equivalent with respect to $\vdash^{-}$.
While for example $- (A\rightarrow B)$ and $A \wedge - B $ are interderivable with respect to $\vdash^{+}$, this does not hold for $\vdash^{-}$.
Fortunately, the answer is that with the definition above we indeed get uniqueness to within synonymy because the following holds in \texttt{2Int}:
If we have equivalence, i.e. interderivability, of formulas both with respect to $\vdash^{+}$ as well as to $\vdash^{-}$, then it is guaranteed that these formulas are also replaceable in any more complex formula, i.e. then it is guaranteed that they are synonymous (for the proof cf. \citealp{Wansing2016a}).
So the upshot of this definition is that we do not only get uniqueness to within equivalence but even uniqueness to within synonymy, without the need to consider compound formulas.

\section{Conclusion}
It has been made clear in other works that there are several features in logical systems which may cause problems for the claim that the connectives are uniquely characterized by the rules of that system.
In this paper I examined the specific problem that occurs in a bilateralist setting in which we have two consequence relations, one for provability and one for dual provability.
The refinements that are needed in such a setting differ from the ones that have been detected so far.
In our specified case we also need to require that the interderivability of the formulas containing the connective is satisfied \emph{for both consequence relations}.
In other bilateral systems the specific formulation of what we require for uniqueness may differ, but in one way or another we will always need a requirement which holds not only for the context of verification (or assertion, or provability), but also for the context of falsification (or denial, or dual provability).

\bibliographystyle{apacite}
\bibliography{ReferencesLogicaSubmission}
\end{document}